\documentclass[a4paper,fleqn,usenatbib]{mnras}

\usepackage{newtxtext,newtxmath}

\usepackage[T1]{fontenc}
\usepackage{ae,aecompl}
\usepackage{xspace}
\usepackage{times}
\usepackage{amsmath} 
\usepackage{amssymb} 
\usepackage{graphicx} 
\usepackage{indentfirst} 
\usepackage{pdflscape} 
\usepackage{placeins} 
\usepackage{nicefrac} 
\usepackage{afterpage} 
\usepackage{rotating} 
\usepackage{booktabs} 
\usepackage{rotating} 
\usepackage{color} 
\usepackage{caption} 
 \usepackage{xprintlen}
 \usepackage[T1]{fontenc}
\usepackage{ae,aecompl}

\usepackage{graphicx}	
\usepackage{amsmath}	
\usepackage{amssymb}	

 \newcommand{\nab}{\mbox{\boldmath$\nabla$}}
 \newcommand{\vel}{\mbox{\boldmath$v$}}
 \newcommand{\kms}{\mbox{\,km\ s${^{-1}}$}}

\newcommand{\bsn}{\texttt{ALL\_Efb\_e001\_5ESN}\space}

\hypersetup{draft} 

\title[Observing the simulated CGM]{Observing the circumgalactic medium of simulated galaxies through synthetic absorption spectra}

\author[Liang, Kravtsov, Agertz]{
Cameron J. Liang$^{1}$\thanks{E-mail:jwliang@oddjob.uchicago.edu}\thanks{NASA Earth \& Space Science Fellow}
Andrey V. Kravtsov$^{1,2}$, Oscar Agertz$^{3}$\\
\\
$^{1}$Department of Astronomy \& Astrophysics, and Kavli Institute for Cosmological Physics, University of Chicago, Chicago IL 60637 \\
$^{2}$Enrico Fermi Institute, The University of Chicago, Chicago, IL 60637, USA \\
$^{3}$Lund Observatory, Department of Astronomy and Theoretical Physics, Box 43, SE-221 00, Lund, Sweden}

\date{Accepted XXX. Received YYY; in original form ZZZ}

\pubyear{2018}

\begin{document}
\label{firstpage}
\pagerange{\pageref{firstpage}--\pageref{lastpage}}
\maketitle

\begin{abstract}
We explore the multiphase structure of the circumgalactic medium (CGM) probed by synthetic spectra through a cosmological zoom-in galaxy formation simulation. We employ a Bayesian method for modelling a combination of absorption lines to derive physical properties of absorbers with a formal treatment of detections, including saturated systems, and non-detections in a uniform manner. We find that in the lines of sight passing through localized density structures, absorption lines of low, intermediate and high ions are present in the spectrum and overlap in velocity space. Low, intermediate and high ions can be combined to derive the mass-weighted properties of a density-varying peak, although the ions are not co-spatial within the structure. By contrast, lines of sight that go through the hot halo only exhibit detectable H\,I and high ions. In such lines of sight, the absorption lines are typically broad due to the complex velocity fields across the entire halo. We show that the derived gas density, temperature, and metallicity match closely the corresponding H\,I mass-weighted averages along the LOS.   We also show that when the data quality allows, our Bayesian technique allows one to recover the underlying physical properties of LOS by incorporating both detections and non-detections. It is especially useful to include non-detections, of species such as  N\,V or Ne\,VIII, when the number of detections of strong absorbers, such as HI and OVI, is smaller  than the number of model parameters (density, temperature, and metallicity).
\end{abstract}

\begin{keywords}
galaxy evolution -- galactic haloes -- circumgalactic medium
\end{keywords}



\section{Introduction}

Due to the low surface brightness of the circumgalactic medium (CGM), nearly all of the observational studies \citep[with few exceptions, e.g., ][]{Martin2014, Cantalupo2014} rely on analyses of absorption spectra from background light sources (e.g., quasars). 
We derive various empirical constraints from mapping the CGM using background quasars absorption spectroscopy.  For example, the abundance of cold/warm absorbers exponentially drops off with increasing distance away from the host galaxies \citep{Nielsen2013, LiangChen2014, Bordoloi2014, Borthakur2016}.  From the spatial profiles and ionization modelling of absorption lines, some researchers derive the total mass and metallicity of the cold phase of the CGM \citep{Lehner2013, Werk2014, Wotta2016, Stern2016, Prochaska2017}.

At the same time, a large set of simulation studies is compared with the constraints of the CGM properties from observations to improve our theoretical understanding. These include the CGM mass budget in different phases \citep{Suresh2015}, the sensitivity of column density profiles and covering fraction of ions with models of star formation and feedback processes \citep{Shen2012, Stinson2012, Hummels2013, FaucherGiguere2015, Liang2016}, and also the metallicity distribution in the CGM \citep{Hafen2016}. 

Just as observers model absorption lines for column density and subsequent ionization analysis for the absorbing gas, simulators should create synthetic absorption spectra to make comparisons more direct \citep[e.g.,][]{OppenheimerDave2006, Bird2015, Churchill2015, Hummels2016}.  In addition, there is likely more information than we currently extract from the observed spectra. Analyzing synthetic spectra generated from simulated gaseous halos with known properties may reveal fruitful insights into how well this inversion process works. Therefore, we develop a pipeline that realistically `observe' the CGM of simulated galaxies by generating synthetic absorption spectra through random lines of sight (LOS).  We study the multiphase structure of the CGM with ionization modelling by combing knowledge about underlying properties from our simulations and the absorption features from the resulting synthetic spectra.

It is worthwhile to note that the spatial resolution of halo gas in the typical cosmological state-of-the-art zoom-in simulations of the kind we use here is far from resolving the parsec scales that may be characteristic of at least some absorbers.  There are theoretical arguments that CGM may be subject to thermal instabilities, which can result in multiphase medium with small-scale structure unresolved in such simulations \citep[e.g.,][]{FallRees1985,BegelmanFabian1990, McCourt2016}. Simulations of small-scale regions focusing on disruption of macroscopic cold/warm regions in a hot halo also show the emergence of very small-scale structures in gas distribution \citep[e.g.,][Liang \& Remming, in prep]{McCourt2016,SchneiderRobertson2017}.  At the same time, some of the current galaxy formation simulations come close to reproducing column density profiles and covering fractions of some of the ionic species \citep{Stinson2012,Ford2016,Oppenheimer2016,Oppenheimer2017,Liang2016}. For example, one of our simulations with boosted feedback ($E_{\rm fb} = 5E_{\rm SN}$), the \bsn run \citep{AgertzKravtsov2015,AgertzKravtsov2016}, reproduces the observed column density profiles of all low and intermediate ions; but it marginally underproduces O\,VI abundance. The relative contribution of small-scale ($\sim$ pc) features to the overall column densities of various absorbers is still under debate. However, given the fact that simulations are producing column densities close to observations, it is reasonable to employ them in tests of the kind presented in this paper.  

In this paper, we focus on the spatial distribution and physical properties, such as temperature, density, and metallicity of the CGM gas in our simulations and how well these can be derived from modelling their signatures in the absorption spectrum. We go beyond what is usually done by developing a Bayesian approach using full probability distribution of column densities of ions. This allows us to treat detections and non-detections in a uniform manner. The paper is organized as follows. In section 2, we describe the simulations and the methodology of generating synthetic absorption spectra. We discuss the multiphase structure, the origins of absorbers such as cold/warm ``clouds" and hot halo gas as seen from synthetic spectra and our simulations. In addition, we explore the effects of various ionization models on the abundances of commonly observed ions. In section 3, we describe a new methodology for deriving physical properties of absorbers, assuming  ionization mechanism is known, that include proper treatment of upper and lower limits of column densities. Finally, we discuss and summarize our findings in section 4.

\section{Simulations and Synthetic Absorption Spectrum}

\subsection{Simulations, star formation, and feedback models}


In this study we use ``zoom-in'' simulation of $\sim L_\star$ galaxy  presented in \cite{AgertzKravtsov2015,AgertzKravtsov2016}. The simulation was run with the Adaptive Mesh Refinement (AMR) code {\tt RAMSES} \citep{Teyssier2002} in the $\Lambda$CDM cosmology with $\Omega_{\Lambda} = 0.73$, $\Omega_{\rm m} = 0.27$, $\Omega_{\rm b} = 0.045$, $\sigma_8 = 0.8$ and $H_0 = 70\, \kms \rm Mpc^{-1}$ \citep{Komatsu2009}.  The mass of the dark matter particle in the high-resolution ``zoom-in'' region is  $m_{\rm DM} = 3.2 \times 10^5 M_{\odot}$, while the peak spatial resolution in the high-density ISM regions is $\Delta x \approx 75$ pc.  The CGM we analyze here is re-binned from the AMR grid at 683 pc resolution everywhere in the halo, typical of current generation state-of-the-art simulations since refinement criteria focus on the densest regime.  At $z =0$, the galaxy has a stellar mass of $M_* \approx 2\times10^{10} M_{\odot}$ and is forming stars with a rate $\approx 2 M_{\odot }\rm{yr}^{-1}$ in a dark matter halo $M_{\rm vir} \approx 10^{12} M_{\odot}$ ($R_{\rm vir}\approx 260$ kpc). 

The specific simulation we use in this study is one of the re-simulations of galaxies started from the same initial conditions, but varying parameters and implementations of star formation and stellar feedback, details of which are presented in \cite{AgertzKravtsov2015,AgertzKravtsov2016}. In \cite{Liang2016}, we have compared the CGM around galaxies in these different re-simulations to available observations of low-, intermediate-, and high-ionization absorbers and showed that one simulation, labeled \texttt{ALL\_Efb\_e001\_5ESN}, has incidence and covering fractions of absorbers closest to observations. We, therefore, choose this simulation (at $z = 0$) as the basis for tests presented in this study.  

The star formation in this simulation was parameterized as:
\begin{equation} 
\dot{\rho}_\star = f_{\rm{H_2}} \frac{\rho_{\rm{g}}}{t_{\rm{ff}}} \epsilon_{\rm{ff}}.  
\end{equation} 
where the star formation rate $\dot{\rho}_\star$ depends on the molecular hydrogen fraction $f_{\rm{H_2}}$ \citep[computed using model by][]{Krumholz2009}, the local gas density $\rho_{\rm{g}}$ and corresponding free fall time $t_{\rm ff}$, and finally the star formation efficiency per free fall time $\epsilon_{\rm{ff}}$. In the simulation we use here $\epsilon_{\rm ff}=0.01$ was adopted. 

The simulation implements feedback recipe that accounts for radiation pressure, stellar winds, and energy and momentum injection by supernovae. 
In the simulation we use in this study, each supernova is assumed to release $5\times 10^{51}$ ergs of energy, a value that is set higher than the traditionally adopted values to increase the effects of feedback. 
In addition, a dual-energy scheme is used in which a fraction of the energy released by supernovae is stored and evolved as a separate variable to model  the unresolved turbulence and thermal pressure of hot gas generated by supernovae.  This energy density, $E_{\rm fb}$, introduces a new form of pressure, $P_{\rm fb} = (\gamma -1) E_{\rm fb}$, and is evolved with its own equation similar to that of the internal energy, but with a dissipation time scale ($t_{\rm dis})$ of 10 Myr:  
\begin{equation}
\frac{\partial}{\partial t}(E_{\rm fb})+\nab\cdot(E_{\rm fb}\vel_{\rm gas} )=-P_{\rm fb}\nab\cdot\vel_{\rm gas}-\frac{E_{\rm fb}}{t_{\rm dis}} \label{eqn:EfbEqn}
\end{equation}
The dissipation time scale is chosen to match values similar to a few sound crossing times of typical giant molecular clouds (GMC).  Such additional energy variable regulates star formation by controlling the pressure within the ISM and thickness of star-forming disk \citep[see][for details]{Agertz2013}.

\subsection{Synthetic absorption spectra of the simulated CGM} 
\begin{figure*}
\begin{center}
\includegraphics[scale=0.5]{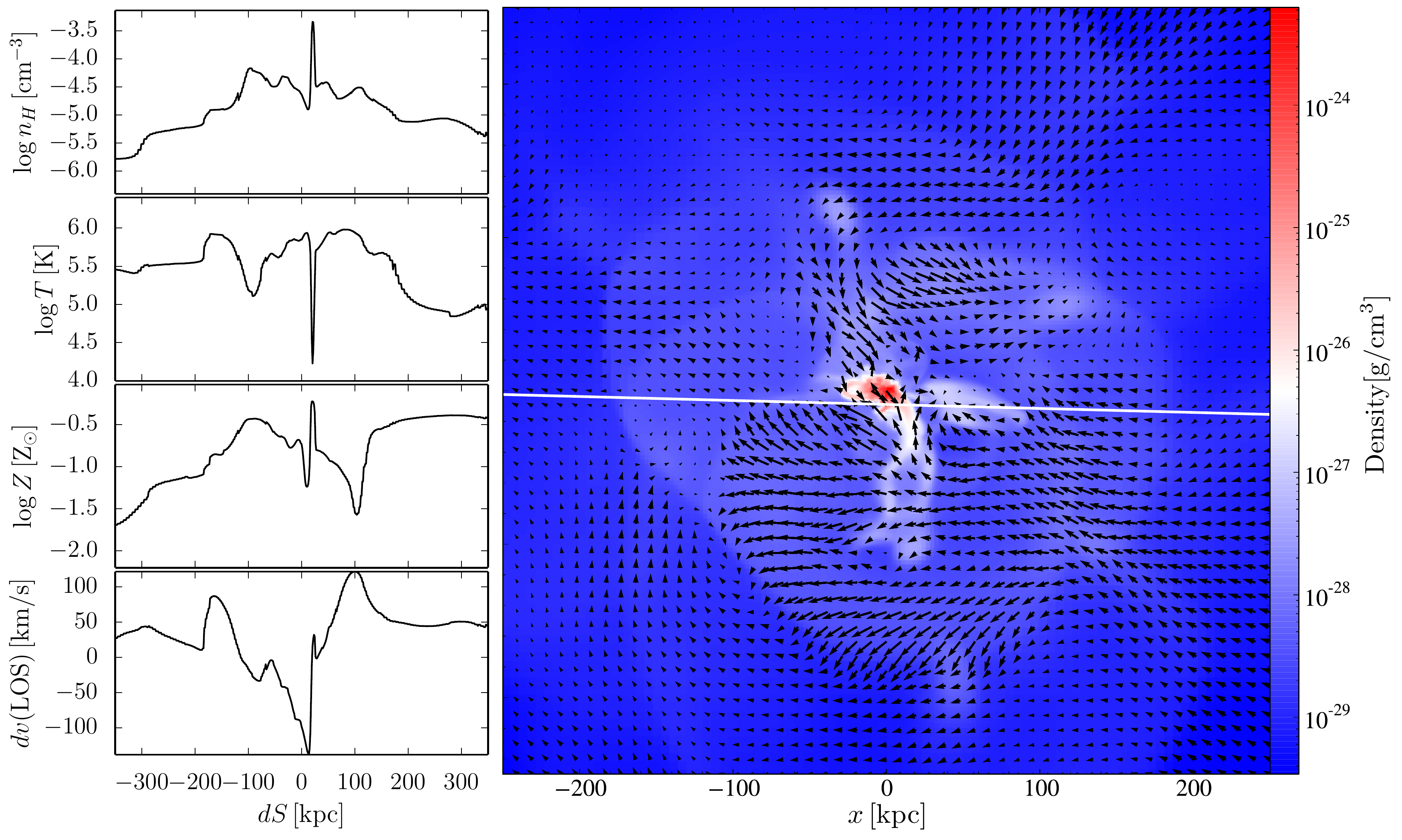}
\caption{Example of a spectrum along a line of sight (LOS) sampling the properties of the CGM around the simulated galaxy. The LOS (white line) through the density (color map) and velocity (vector) field are shown in the right panel, while the left panels show the one-dimensional profiles of the density, temperature, metallicity, and velocity (relative to the galaxy systemic velocity) of the gas the LOS passes through.  \label{fig:los_map}}
\end{center}
\end{figure*}

The synthetic spectra are computed with our custom code that converts physical properties of absorbing gas (such as density, temperature, and metallicity) into absorption lines, as described below. 

First, we compute the number densities of ions from random LOS through the simulations.  Briefly, the number density of an element X (e.g., C, N, O) at state $j$ is computed as: 

\begin{equation} n_{X_j} (n_H, T, Z) = f_{X_j} (T, n_H) f_X(Z)  n_H,\end{equation}

where $f_X(Z) = n_X / n_H$ is the fraction of the element $X$ relative to hydrogen assuming solar pattern. The ionization fraction, $f_{X_j} (T, n_H) = n_{X_j} / n_X$, is computed with both photo- and collisional ionization using the \texttt{CLOUDY} code \citep{Ferland2013}. We assume that each simulation cell $i$ represents an isothermal cloud with temperature $T$ and constant hydrogen density ${n_H}$.

The integrated normalized spectrum flux at wavelength location $\lambda$ is thus computed assuming optically thin photoionization from the \cite{HaardtMadau2012} ultra-violet background (UVB) as 
\begin{equation} 
F(\lambda) = \prod_i \exp[-\tau_i (\lambda)], 
\end{equation}
where $i$ is the index running through computational cells along a given LOS and the optical depth $\tau_i$ of each cell $i$ is given by:
\begin{equation} 
\tau_i (\lambda | N_i, b_i, z_i)  = N_i \frac{\pi e^{2}}{m_e c} f_{\rm osc} \Phi (\lambda | b_i,z_i).
\end{equation}
Here $N_i$ is the column density of the corresponding ion $N_{X_j} = n_{X_j} dl$ computing using ion number density in each cell and its size $dl$, $b_i$ is the Voigt width parameter set by the temperature and velocity distribution of gas along LOS, and $z$ is the redshift of the cell relative to the assumed systemic redshift of the galaxy, which includes both the peculiar gas motion and the Hubble flow contribution, $v_{\rm H} = H_z r$, where $H_z$ is the Hubble constant at the systemic redshift $z$. Also, $e$, $m_e$, $c$ and $f_{\rm osc}$ are the relevant constants for the ion: the electron charge, electron mass, speed of light and oscillator strength of the transition between energy levels of the atom, respectively. Finally,  $\Phi(\lambda | b_i,z_i)$ is the Voigt profile, which we computed using the real part of the  Faddeeva function using the {\tt wofz} function of the \texttt{SciPy} package.\footnote{\url{https://docs.scipy.org/doc/scipy/reference/generated/scipy.special.wofz.html}\\ 
\url{http://ab-initio.mit.edu/wiki/index.php/Faddeeva_Package}}. 

Figure \ref{fig:los_map} shows an example of a random LOS going through our simulated galaxy in the \bsn run. The density map shows an aerial view of the LOS that goes through some inflowing materials close to the star-forming regions. The LOS records the one-dimensional information shown quantitatively in the left panel of Figure \ref{fig:los_map}, covering a wide range of density, temperature, metallicities and projected velocities. The corresponding ionic abundance and absorption spectra are shown in Figure \ref{fig:mp_spat} and \ref{fig:mp_spec}. These indicate the multiphase nature of the gas along this LOS.  With the tools to generate synthetic spectra for any ions and at any viewing angle of the simulated CGM, one can then systematically study both the cold and hot gas in the gaseous halos, which we will discuss in the following sections.

\subsection{Muti-phase gas viewed from simulations and synthetic spectra}

\begin{figure*}
\begin{center}
\includegraphics[scale=0.5]{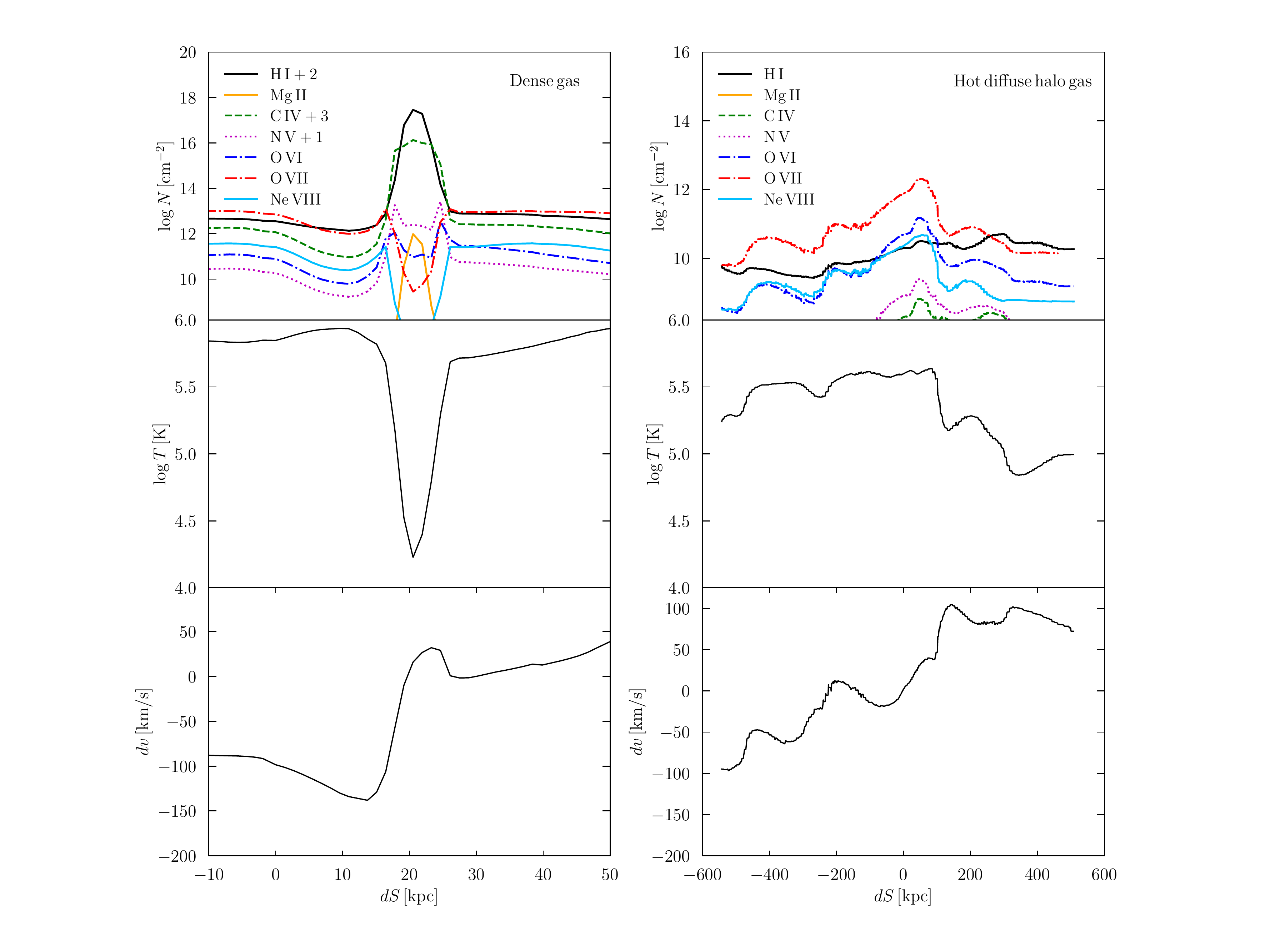}
\caption{One-dimensional profiles of column density, temperature, and velocity as a function of distance from the point of closest approach (minimum distance) to the galaxy for two types of lines of sight (LOS). Note that the top plots of both panels show column density contributed by each pixel $i$ at the location $dS_i$, i.e., $N_i = n_i dS_i$. The total column density is integrated along the LOS, $N  = \sum N_i$. For clarity, various lines representing different ions have been artificially shifted by the value shown in the legends. The left column shows the multiphase structure of the absorbing gas in the LOS shown in Figure \ref{fig:los_map}. This LOS passes through a density peak with a large gradient in gas density and temperature. High ions (e.g., N\,V, O\,VI, O\,VII and Ne\,VIII) preferentially reside on the boundary of the density peak while low ions reside in the core where the temperature is low ($\sim 10^4$ K).  The right panel shows the LOS that passes through the hot and diffuse coronal halo gas near the virial temperature of the halo ($\sim 10^{5.5}$ K) without any dense gas. The abundance of low ions is therefore small.  The temperature and velocity profiles (middle and lower columns) determine the line width of absorbers in the absorption spectra shown in Figure \ref{fig:mp_spec}. \label{fig:mp_spat}}
\end{center}
\end{figure*}

\begin{figure*}
\begin{center}
\includegraphics[scale=0.5]{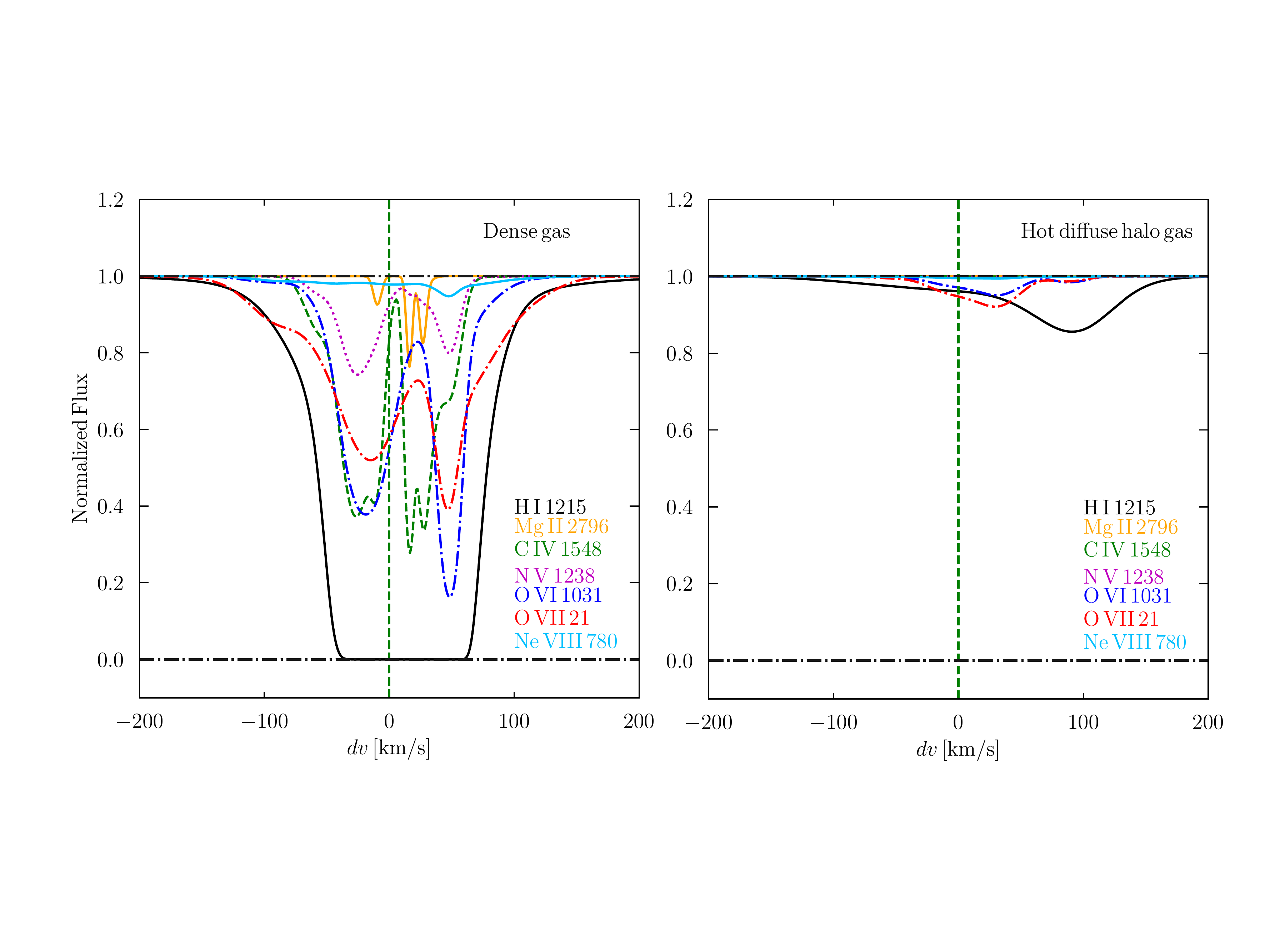}
\caption{The synthetic spectra generated from the two LOS shown in Figure \ref{fig:mp_spat}. Left panel: the LOS that passes through the density peak exhibits strong absorption lines of low ions, as well as high ions (e.g., OVI), reflecting the multiphase structure of the gas along this LOS. Note that a single density peak along this LOS generates three (or four) absorption components due to the strong velocity gradients. However, the low and high ions are approximately kinematically aligned per absorption component. However, the large column density of H\,I and oscillator strength of Ly$\alpha$ lead to a blending of all three absorption components, making the determination of physical parameters difficult, although such blending does not always occur. Higher order Lyman series lines are needed to separate the absorption components. Right panel: the LOS that passes through the diffuse hot halo exhibits only H\,I and O\,VI absorption since low ions cannot survive in the physical condition near the virial temperature of the halo ($\sim 10^{5.5}$ K). We also find detectable O\,VII lines ($\lambda_{\rm rest} =21.6 $\AA) in both cases in our simulations with $\log N \sim 14\,\rm{cm^{-2}}$. \label{fig:mp_spec}}
\end{center}
\end{figure*}

Absorption spectroscopy routinely shows that the CGM is multiphase and dynamic; the evidence for multiphase gas is found in a wide range of galaxies \citep[e.g.,][]{Hennawi2006, Steidel2010, Tumlinson2011,Werk2014,Bordoloi2014,LiangChen2014,Prochaska2014, Johnson2015b,Borthakur2016, Lau2016, Heckman2017}. Multiphase gas is generically expected where cold/warm absorbing gas is moving through a hot halo. In our simulations, the cold/warm gas in the hot halo corresponds to outflows of the ISM gas driven by supernovae feedback and, partly, accretion of gas from the IGM.  In this section, we show how the underlying physical properties of multiphase gas manifest in spectroscopic observations by generating synthetic spectra through LOS in the simulated CGM. 

In Figure \ref{fig:mp_spat}, we show the physical properties of two types of LOS in our simulations. The left panel shows an LOS that goes through a dense absorber with large gradients in density and temperature.  
It is this large range of density and temperature in a localized region that simultaneously give rise to an abundance of low, intermediate and high ions. Naturally, these ions are not co-spatial. The low ions (H\,I and Mg\,II) are produced in the densest core of the absorber whereas the intermediate and high ions (C\,IV, N\,V, O\,VI and Ne\,VIII) are spatially extended in the boundaries where the density is low and the temperature is high. By contrast, the right panel of Figure \ref{fig:mp_spat} shows an LOS that goes through the hot halo itself with characteristic temperature $T_{\rm vir} \approx 10^{5.5} $ K for an $L_*$-like galaxy. For this type of LOS, the physical condition leads to highly ionized gas where the abundance of O\,VI and Ne\,VIII is orders of magnitude higher than low ions.  

Using the formalism outlined in sec 2.2, Figure \ref{fig:mp_spec} shows the corresponding synthetic spectra for the two LOS in velocity space centered at the systemic redshift of the galaxy.  The left panel shows simultaneously the absorption lines from low to high ions, indicating the existence of multiphase gas discussed above. Note that the absorption lines are split into multiple velocity components (roughly three) due to the velocity shear across the absorber.  The number of absorption components depends on both the spatial resolution to resolve individual absorbers and velocity/spectral resolution for their motion.
The alignment of the strongest absorption component between the low and high ions depends on how quickly the velocity varies across the size of the absorber.  The alignment of O\,VI with low ions seen in observations \citep[e.g.,][]{Werk2016} suggests that the velocity field does not vary significantly over a range of sub-kiloparsec based on cloud size estimates. These are inferred from the required low density assuming photoionization modelling in \cite{Stern2016}. 

The right panel of Figure \ref{fig:mp_spec} shows the corresponding absorption spectrum of the hot halo gas.  In this particular LOS, we see detectable broad O\,VI and H\,I lines but not Ne\,VIII lines. In this type of LOS, the absorption profile is typically very broad due to the large variation of the velocity field across the entire halo ($\sim 250$ kpc), although rare exceptions do exists in some particular viewing angles.  In addition, the dominant ionization state of oxygen is O\,VI at $T_{\rm vir} \sim 10^{5.5} $ K.  Our simulations suggest the observed LOS with similar characteristics (i.e., broad O\,VI and no low ions) are likely probing the hot halo gas of galaxies with $T_{\rm vir} \sim 10^{5.5} $ K \citep{Savage2011a, Werk2016}.

Although the integrated column density for Ne\,VIII within the virial radius is $\approx 10^{13} {\rm cm^{-2}}$ \citep[see Fig. 15 in][]{Liang2016}, the optical depth is spread out in velocity space which makes it difficult to detect observationally because it can be easily hidden in the noise and in the uncertainty of continuum placement.  In section 3, we present a Bayesian Voigt profile routine that includes utilities such as a continuum model to account for this uncertainty of continuum placement. We also describe a method of constraining physical properties of the hot halo with a limited number of detections and poor constraints based on the Bayesian approach.

\begin{figure*}
\begin{center}
\includegraphics[scale=0.55]{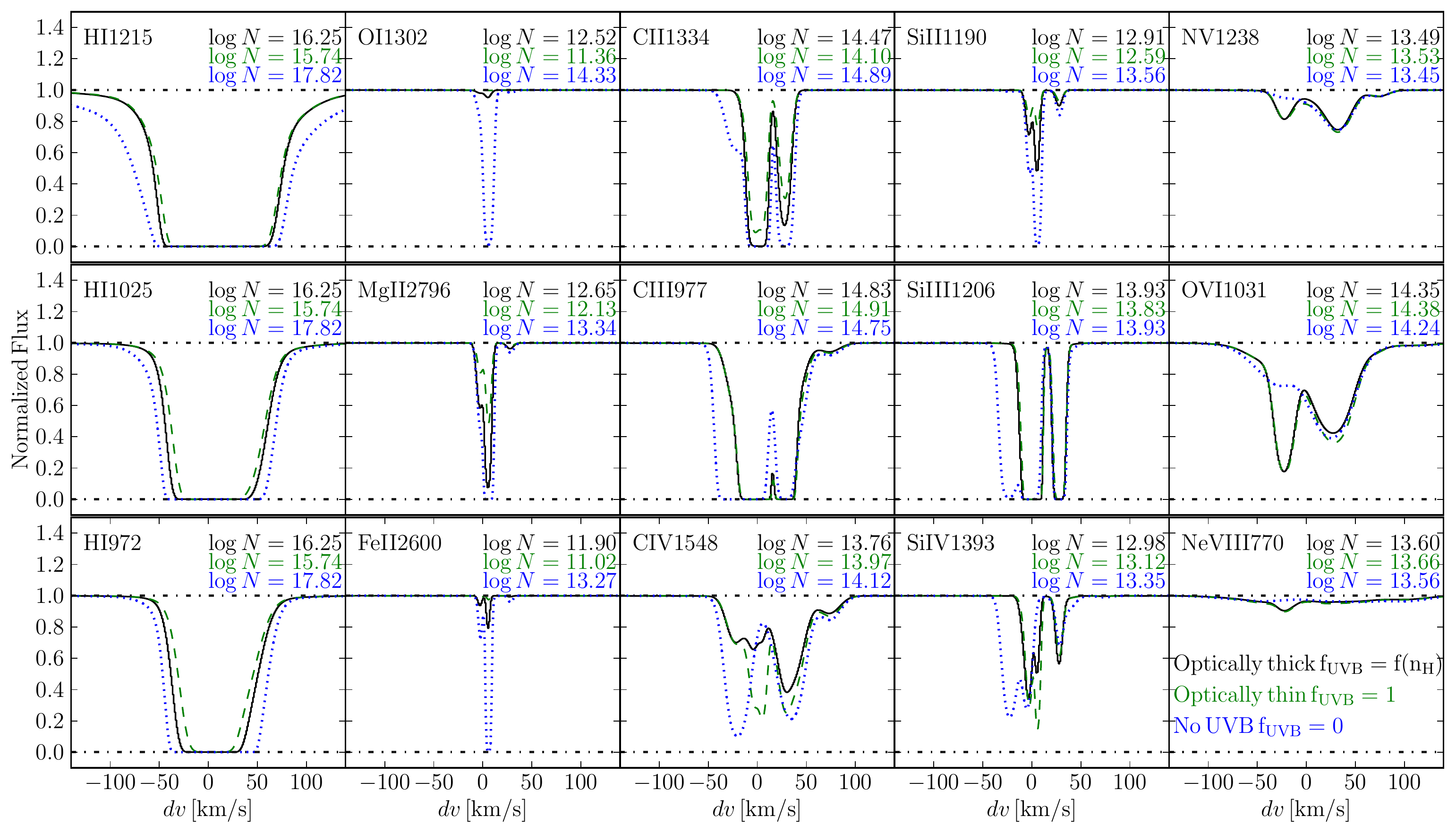}
\caption{This figure shows the same LOS as the left panel in Figure \ref{fig:mp_spec} but with variations of ionization models. This shows how ionization models affect the total column densities of low, intermediate and high ions and their spectral features.  The blue spectrum is computed using an ionization model with photoionization turned off (i.e collisional ionization only). The green spectrum assumes the gas is optically thin with 100\% of the UVB contribution. Lastly, the black curve takes into account self-absorption of the CGM gas using an approximation for the fraction $f_{\rm UVB}$ that reaches the gas as a function of the local density from \protect\citet{Rahmati2013}.  Comparing the model with collisional ionization only ($f_{\rm UVB} = 0$), the optically thin photoionization model ($f_{\rm UVB} = 1$) extinguishes a large fraction of low ions (e.g., H\,I, O\,I, Mg\,II and Fe\,II; blue to green), which results in the change of shape of the absorption profile. Absorption lines of OVI, however, change relatively little (green vs blue profiles), which shows that they are primarily collisionally ionized.  Taking into account CGM self-absorption (black) generally increases the abundance of low ions compared to the optically thin model (green). \label{fig:spec_ion_model} }
\end{center}
\end{figure*}

\begin{figure*}
\begin{center}
\includegraphics[scale=0.65]{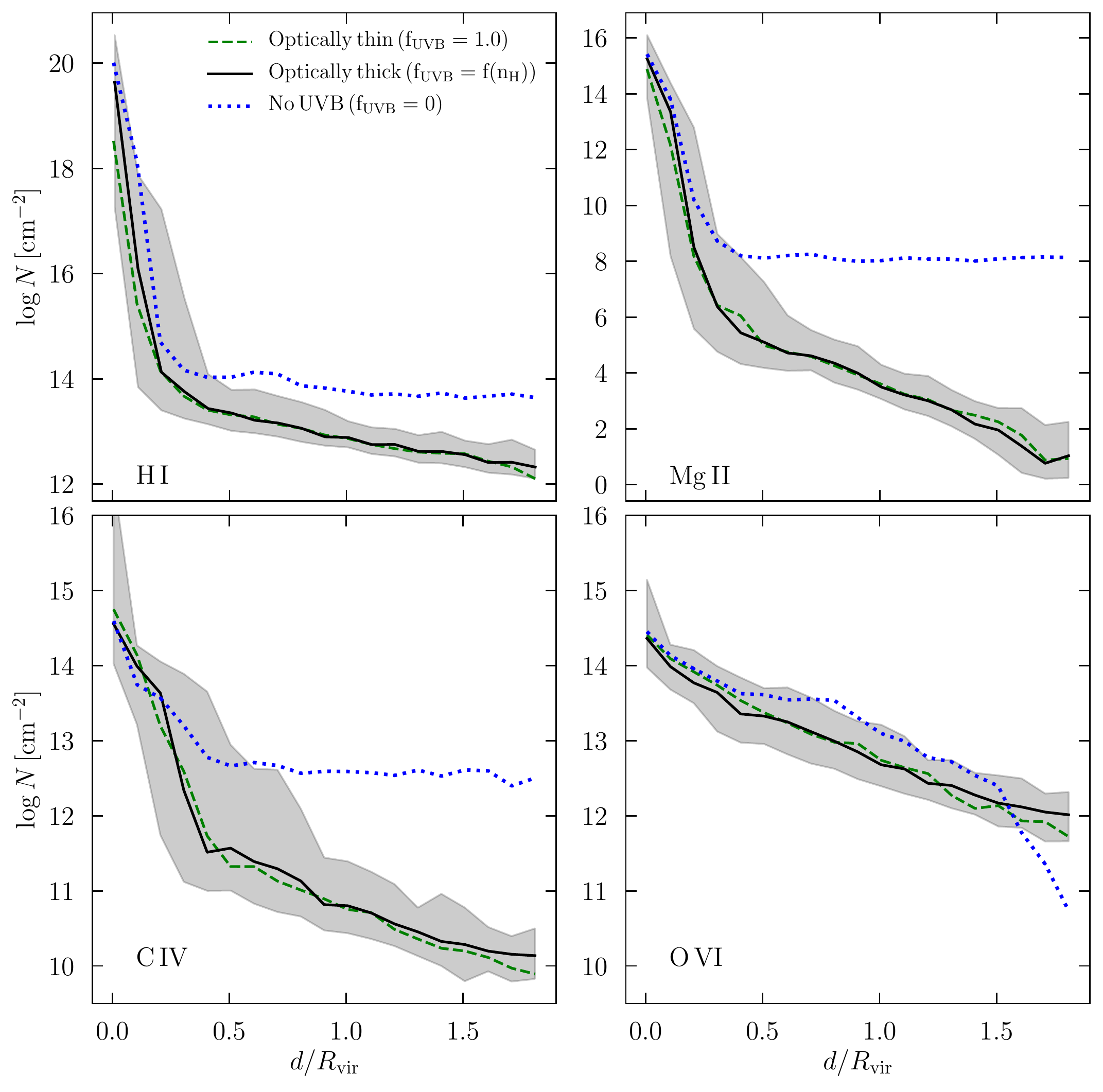}
\caption{Column density profiles of H\,I, Mg\,II, C\,IV and O\,VI computed for the different treatment of ionization radiation (see legend and caption of Figure \ref{fig:spec_ion_model}). The comparison of different ionization models show that photo-ionization from the UVB extinguishes an abundant of low and intermediate ions at distances beyond $\approx 0.2 R_{\rm vir} \approx 65$ kpc. On the other hand, photo-ionization does not have a strong effect on the column density of O\,VI, showing that it is primarily collisionally ionized. The model that takes into account CGM self-absorption generally increases the abundance of low ions. \label{fig:ionpro}}
\end{center}
\end{figure*}

\subsection{Relative contribution of photoionization and collisional ionization}

In Figure \ref{fig:mp_spat} and \ref{fig:mp_spec}, the abundance of ions and their spectral features are computed using an ionization equilibrium model that assumes collisional and photoionization in the optically thin regime. The relative contribution of these ionization processes for different ions in observed absorption systems is not yet completely understood.  It is thus interesting to explore the relative contribution of photo- to collisional ionization using our simulations. The interesting questions include, for example,  the degree to which the abundance of low ions in the core of absorbers can be boosted in the optically thick regime.

To correctly compute the incident ionizing flux, one must in principle perform radiative transfer calculation. Although no radiative transfer was carried out in the simulation, we compute the optical depth approximately in post-processing using a local approximation for the attenuation of the ionizing photons calibrated by \cite{Rahmati2013}. Specifically, we use Eq. (6) to parameterize the ratio of the attenuated photoionization rate ($\Gamma_{\rm Phot}$) to the total background rate ($\Gamma_{\rm UVB}$) as a function of the local density of a computational cell: 

\begin{equation}
\begin{split}
 f_{\rm UVB}(n_{\rm H}) \equiv \frac{\Gamma_{\rm Phot}}{\Gamma_{\rm UVB}} = 0.98 \left[ 1+ \left( \frac{n_{\rm H}}{n_{\rm {H,ssh}}} \right)^{1.64} \right]^{-2.28} \\
 + 0.02\left[ 1+ \frac{n_{\rm H}}{n_{\rm {H,ssh}}}  \right]^{-0.84}
 \end{split}
 \end{equation}

where $n_{\rm {H,ssh}}$ is a free parameter calibrated for effects of self-shielding, which depends on the UVB. For our choice of baseline UVB model \citep[][HM12]{HaardtMadau2012} and our analysis of the simulation snapshot at $z = 0$, we have adopted $n_{\rm {H,ssh}} = 5.1 \times 10^{-4} \rm{cm^{-3}}$ \citep[table 2 in][]{Rahmati2013}. We then run a grid of \texttt{CLOUDY} \citep{Ferland2013} models with varying ratio ($f_{\rm UVB}$), density, and temperature in order to compute the modified ion fraction (e.g., $f_{\rm HI}  = f_{\rm HI}(n_H, T, f_{\rm UVB}$)).  For a given simulation cell with $n_H$ and $T$, the ionization fraction is uniquely determined since $f_{\rm UVB}$ is also a function of $n_H$.

Figure \ref{fig:spec_ion_model} and \ref{fig:ionpro} show the resulting spectra and the overall abundance of ions from three ionization models with varying degree of contribution from photoionization: optically thin ($f_{\rm UVB} = 1$), taking into account CGM self-absorption ($f_{\rm UVB} = f_{\rm UVB}(n_{\rm H})$), and no photoionization ($f_{\rm UVB} = 0$).  Figure \ref{fig:ionpro} shows that taking into account
attenuation of the background by self-absorption of the CGM  does not affect the median column density profiles of ions substantially, as the profiles for $f_{\rm UVB} = 1$ and $f_{\rm UVB} = f_{\rm UVB}(n_{\rm H})$ at larger radii are nearly identical. This is because the covering fraction of dense and cold gas is small. At small radii ($r/R_{\rm vir}\lesssim 0.2R_{\rm vir}$) and high densities, accounting for self-absorption boosts the column densities of low ions by a factor of a few to ten (see the O\,I, Fe\,II, Mg\,II and C\,II panels in Fig \ref{fig:spec_ion_model}). This difference is smaller than the scatter of the profiles for different LOS for this object.  

The OVI profile is not sensitive to the CGM self-absorption corrections because in our simulations OVI absorbing gas \textit{mostly} arises in hotter and less dense collisionally ionized gas.  This is reflected in the relative insensitivity of the OVI column density profile to the presence of UVB radiation. HI, CIV, and Mg II ion column densities, however, are very different for the $f_{\rm UVB} = 0$ case at $r/R_{\rm vir}\gtrsim 0.2R_{\rm vir}$ and thus show significant contribution from photoionization at these radii. Photoionization is significant for low ions for these large radii due to their low ionization potential and the decrease in optically thick gas with increasing radius. This also illustrates that photoionization shapes the exponential form of the column density profiles in Figure \ref{fig:ionpro}, as discussed in \cite{Liang2016}. 

\section{A Bayesian Approach for Extracting Physical Properties of the CGM}

\begin{figure}
\begin{center}
\includegraphics[scale=0.35]{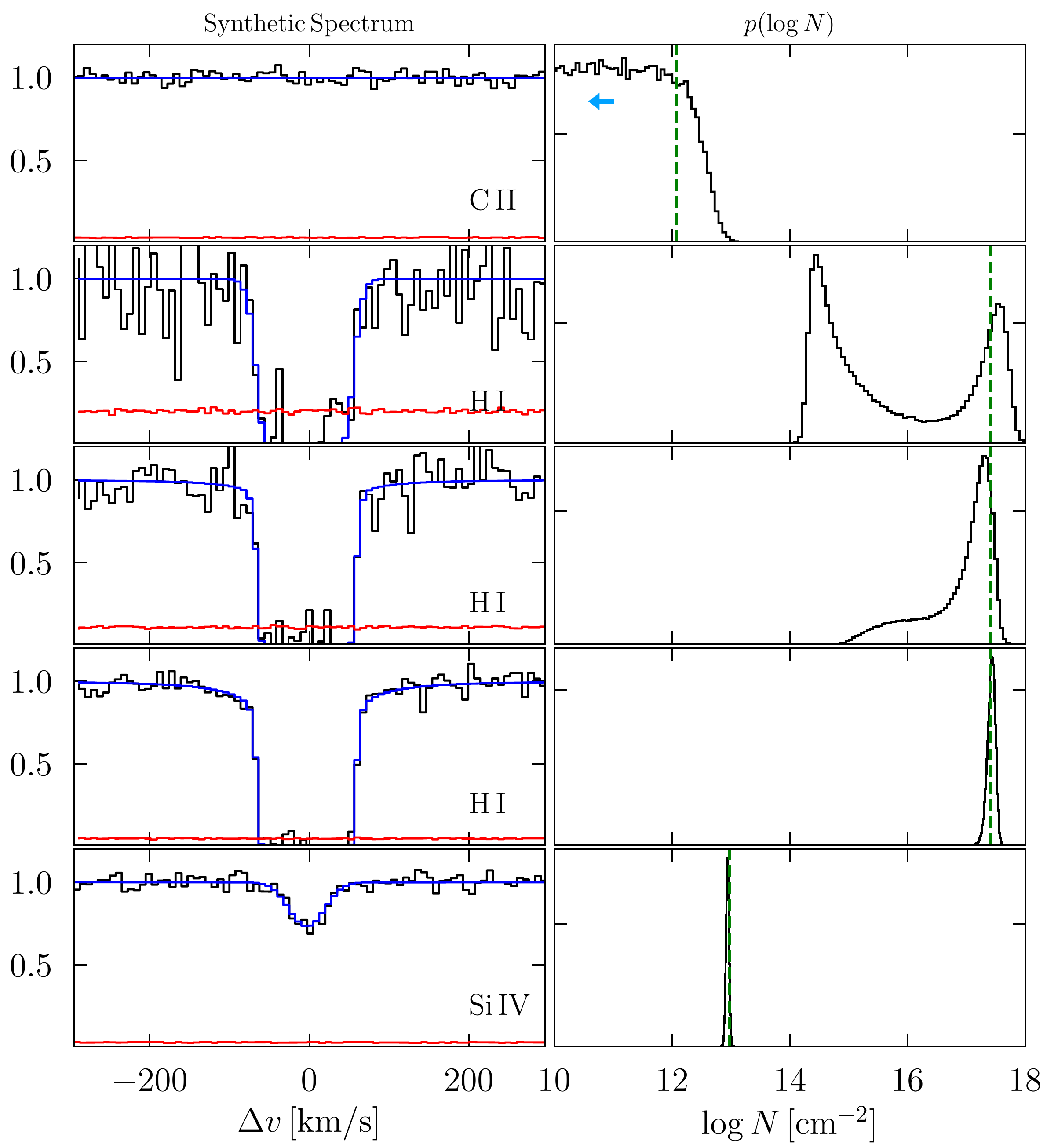}
\caption{Each row shows synthetic spectrum (with continuum $C(\Delta v)$ = 1) around a particular absorption line derived from an LOS through simulation along with the best fit model Voigt profile resulting from our Bayesian fitting (left panel) and the derived posterior probability distribution of the column density of the corresponding. The true column densities are indicated in the right panel by the green vertical dashed lines. We have selected five representative examples of absorption lines that show different degrees to which the lines were detected: from a non-detection (top row) to high signal-to-noise detection (two bottom rows). The same saturated H\,I line is repeated three times in the middle panels with varying $S/N$ (= 5, 10, 20) showing that the resulting fit can be non-Gaussian in low $S/N$ spectra due to the degeneracy between $N$ and $b$-parameter. For the non-detection/upper limit constraints in C\,II $\lambda 1334$, the lack of decrement in the flux implies that column density of C\,II greater than $\sim 10^{13}\,\rm{cm^{-2}}$ is disfavored by the data. In contrast, the Si\,IV is well measured with a narrow Gaussian column density PDF. These probability distributions can then be used for ionization modelling even though C\,II line is not formally ``detected" or that the H\,I line is saturated. \label{fig:vpfit}}
\end{center}
\end{figure}

\begin{figure*}
\begin{center}
\includegraphics[scale=0.65]{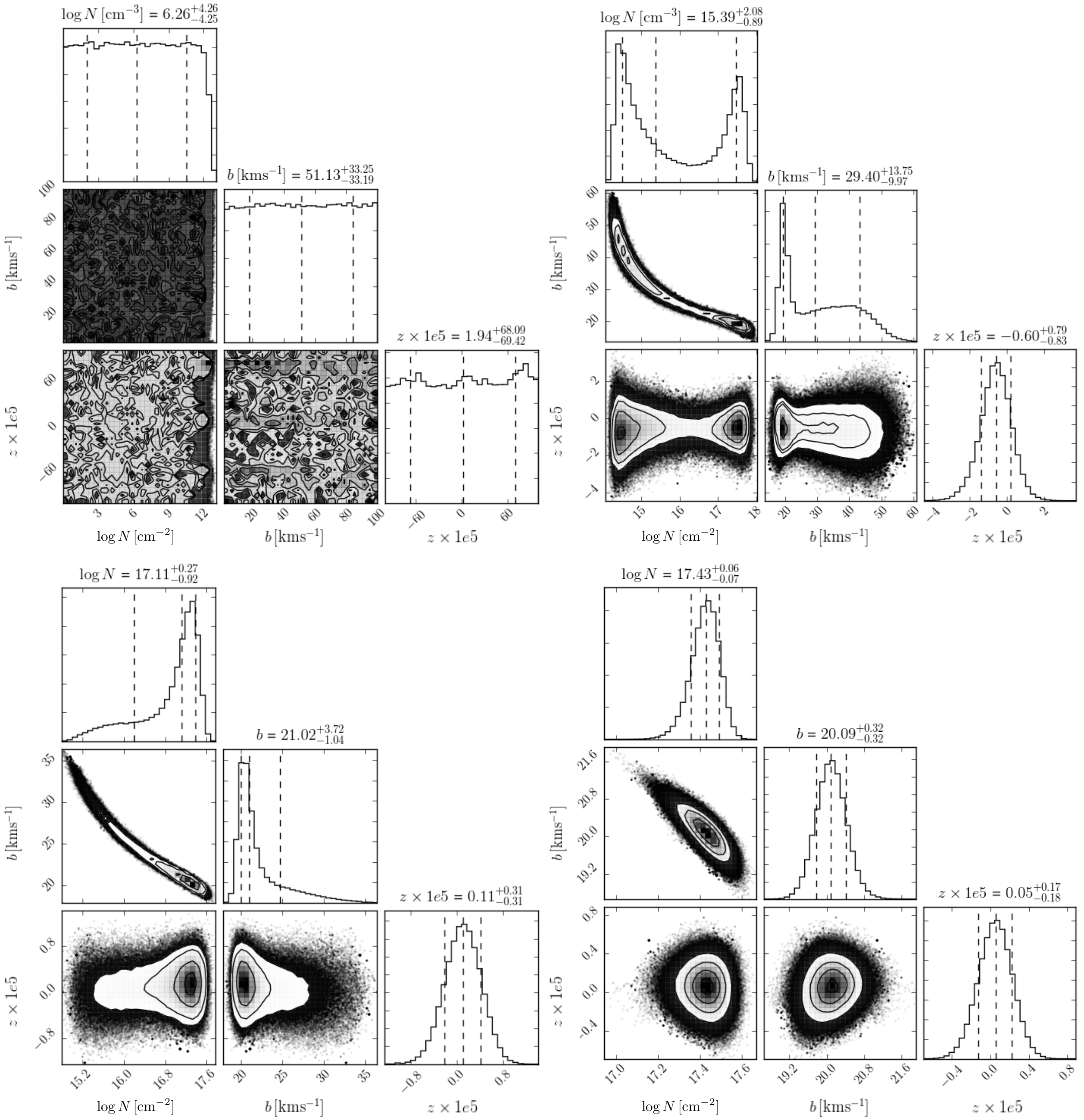}
\caption{These four corner plots show the joint posterior distributions corresponding to absorption lines in Figure \ref{fig:vpfit}.  The top left panel shows the posteriors of the non-detected C\,II line. The joint probability distributions show noisy fluctuation without a preferred location in the parameter space. The marginalized probability of column density drops precipitously for $\log N/cm^{-2} > 12$, where high column density is strongly disfavored by the data.  The other three panels show the posteriors of the saturated H\,I line with corresponding S/N of 5, 10 and 20. The elongated shape in the joint probability shows the degeneracy between $b$ and $N$ in a saturated regime. As expected, the constraints improve as S/N of the spectra increases. The corner plots are made using publicly available code \texttt{corner.py} \citep{corner}. \label{fig:vp_corner}}
\end{center}
\end{figure*}

\begin{figure*}
\begin{center}
\includegraphics[scale=0.5]{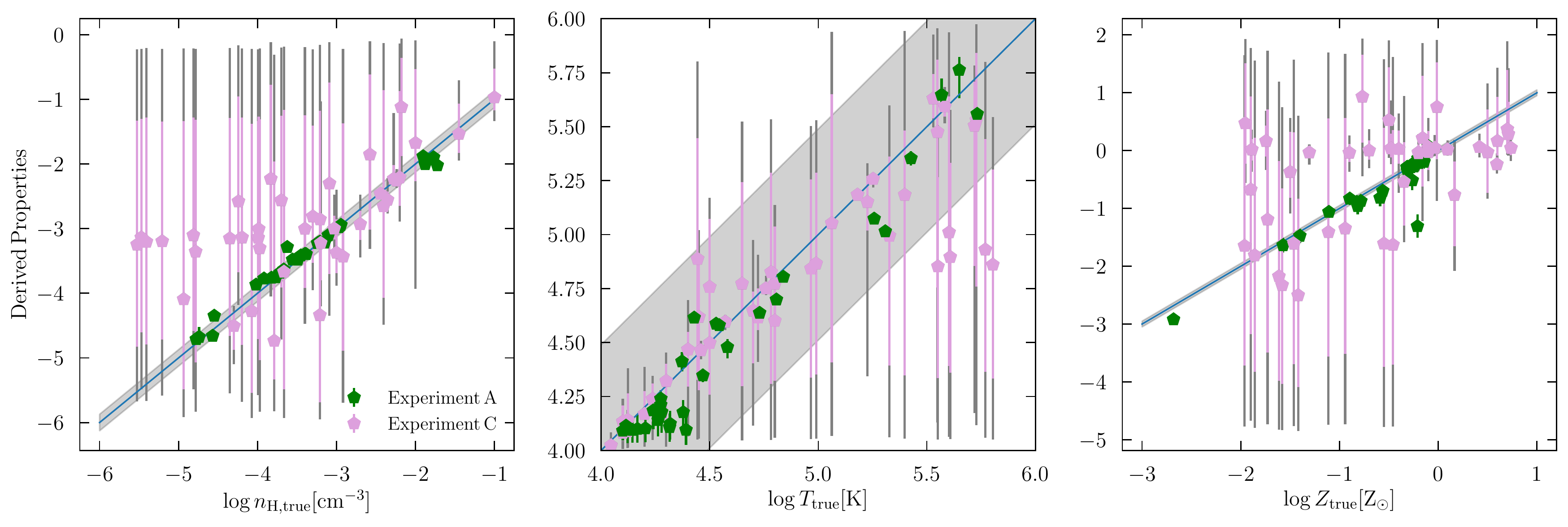}
\caption{Comparison between derived properties and the ``true" HI mass-weighted averages of each quantity along corresponding LOS in simulations. Experiment A (green points) represent values derived using the integrated column density along the entire LOS, which may include absorption from multiple absorbers. The column densities in experiment A have uncertainties of 0.1 dex.  Generally, the derived values are close to the HI-weighted properties along the LOS. Experiment B (not shown) shows similar results as A . The gray bands show the typical intrinsic scatter along these LOS.  The pink points in experiment C show how well one can retrieve the physical properties of absorbers when column densities constraints are poor (see the upper limit constraint of C\,II in Fig \ref{fig:vpfit} for an example). The pink error bars show the ranges that enclose 68\% and the gray error bars enclose 99\% of probability of the posterior distributions. Naturally, the constraints of the derived parameters are poorer due to the lack of constraining power from upper limits PDF. Nevertheless, this demonstrates a bias-free, formally correct way to retrieve physical properties of the CGM when there's limited information available.   \label{fig:props_comp} }
\end{center}
\end{figure*}

\begin{figure}
\begin{center}
\includegraphics[scale=0.62]{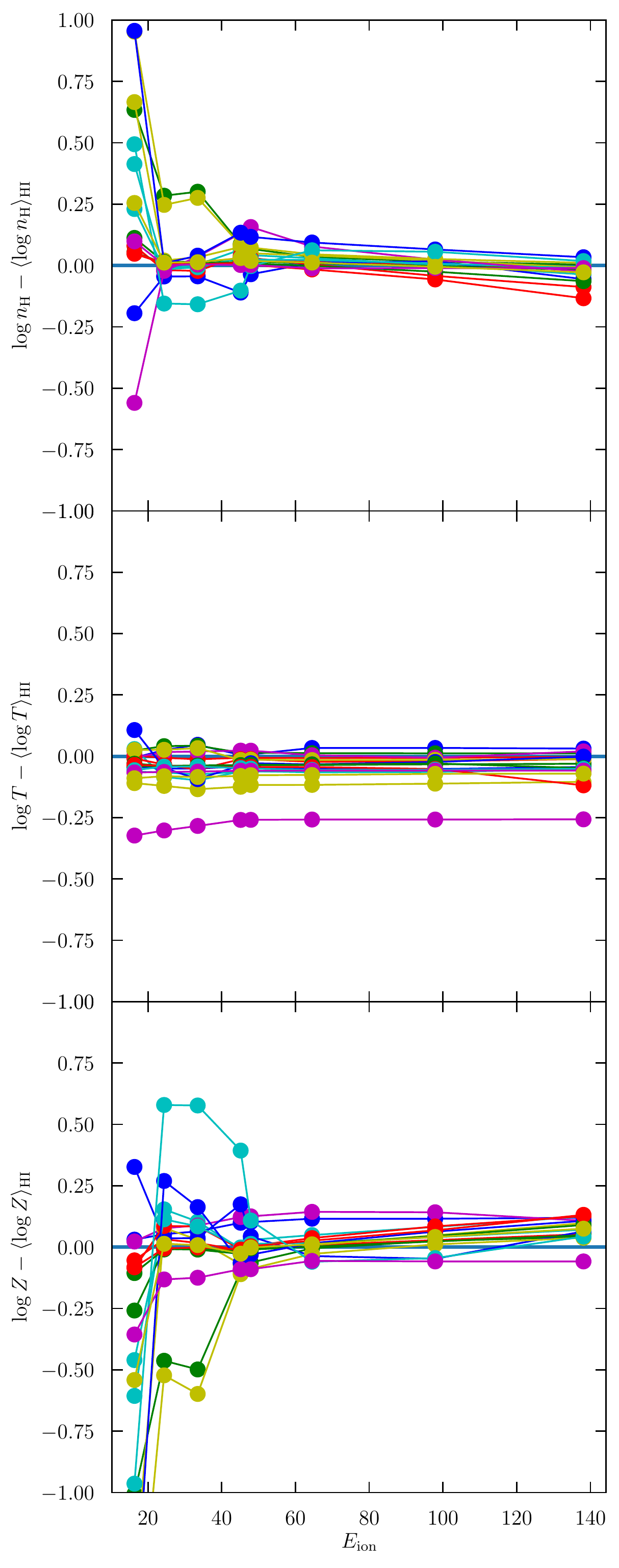}
\caption{Experiment B: comparison of how well ionization modelling can retrieve the H\,I density weighted physical properties by including a set of ions with increasing ionization energy. Each line connected by color points represent a single LOS. The color points at a given $E_{\rm ion}$ means that the ionization modelling includes ions \textit{up to} the specified $E_{\rm ion}$. For example, points at $E_{\rm ion} = 16.345$ eV means H\,I ($E_{\rm ion} = $13.598 eV), O\,I (13.618), Mg\,II (15.035) and Si\,II (16.345) are included in the ionization modelling. The typical error bars are larger with a low number of ions (not shown), and all derived properties are roughly consistent with the input values. Surprisingly, the derived physical properties approach the H\,I mass-weighted values by including higher ions.\label{fig:ionfitB} }
\end{center}
\end{figure}

\begin{figure}
\begin{center}
\includegraphics[scale=0.37]{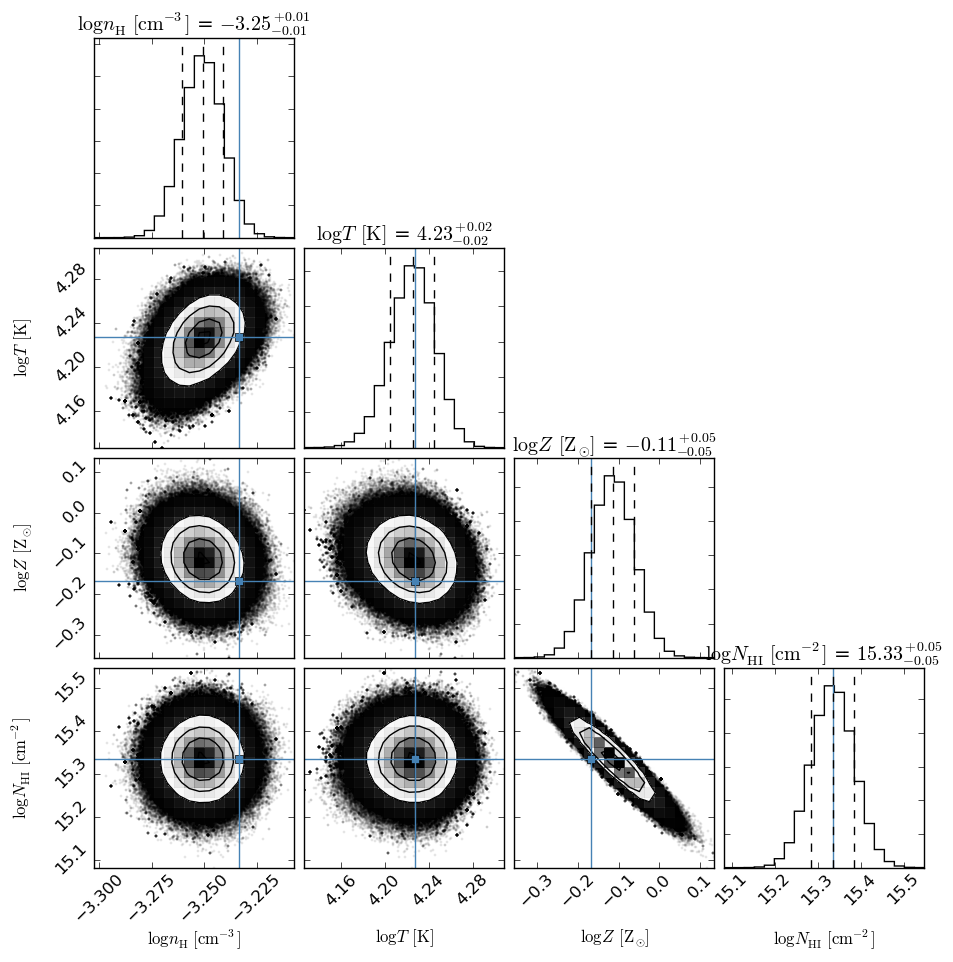}
\caption{An example of posterior PDFs of the parameters in experiment B, in which all low to high ions with $E_{\rm ion} \le 140$ eV (up to O\,VI, with $\log N_{\rm OVI} /\rm{cm^{-2}}= 14.06$) are included. As shown, the derived properties are fairly consistent with the underlying H\,I-weighted values. 
\label{fig:ionfitB3} }
\end{center}
\end{figure}

\begin{figure*}
\begin{center}
\includegraphics[scale=0.7]{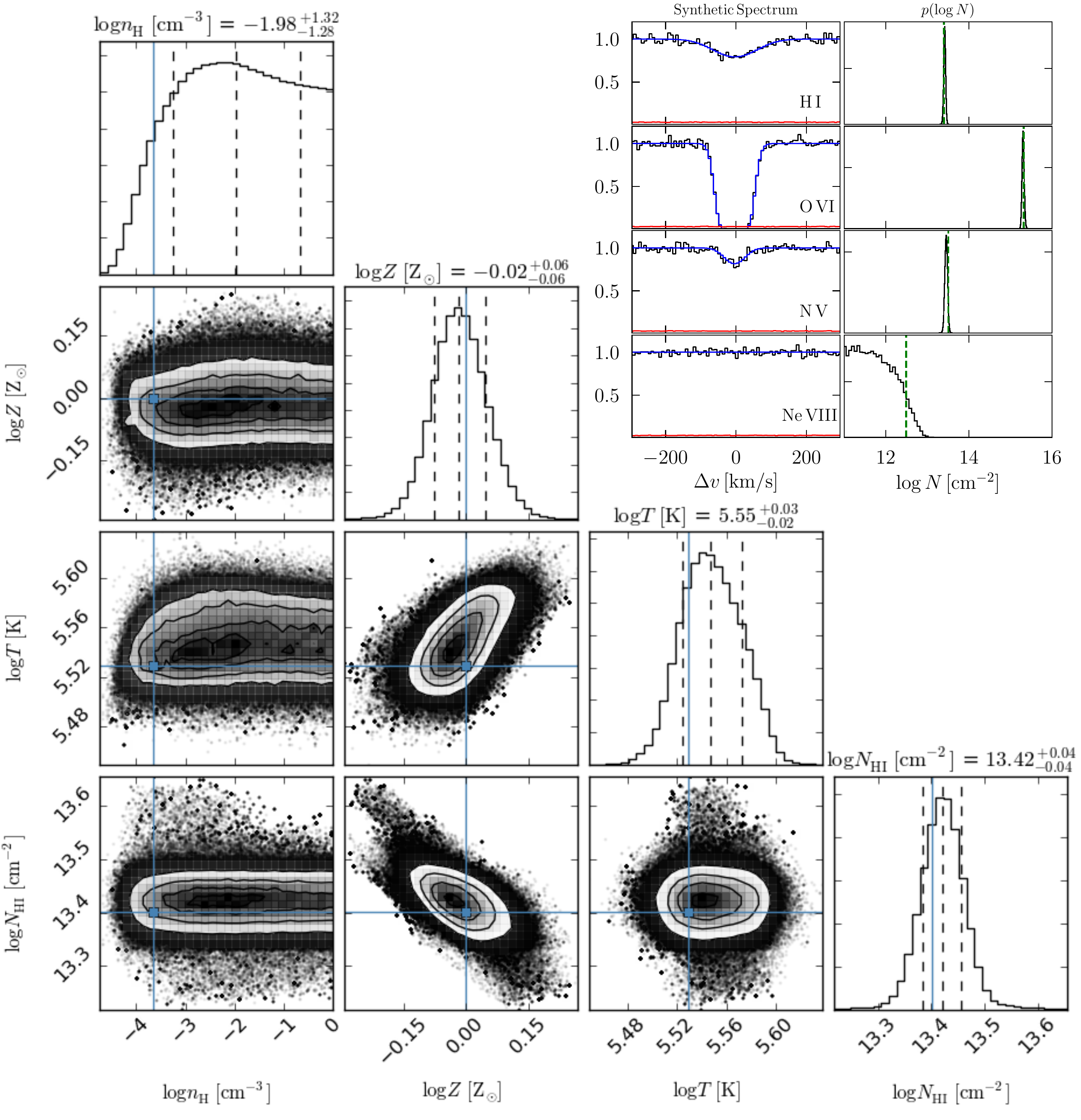}
\caption{MCMC sampled posterior distributions of best-fit parameters of an ionization model (including photo- and collisional ionization) derived from constraints of column density in the form of non-Gaussian probability distributions including ``non-detected" metal lines. In this example,  column density constraints come from H\,I, O\,VI, N\,V and Ne\,VIII, where Ne\,VIII is a ``non-detection" similar to C\,II constraints in Figure \ref{fig:vpfit}. The corresponding input mock spectrum and derived column density PDF is shown top right. Although with the poor column density constraints with Ne\,VIII, this figure shows one can formally recover the true parameters (shown by blue crosses) in this mock observation to the degree that the data allow. If N\,V becomes an upper limit when S/N of the spectrum is poorer, the corresponding constraints of temperature and metallicity will be poorer. Explicit priors can also be set to limit the wide range of allowed density based on our physical understanding of O\,VI-producing gas.
Note that the probability distribution of H\,I column density from Voigt profile fit acts as a prior for the ionization modelling. In other words, the posterior of H\,I shown here is a re-sampled version of the posterior sampled from Voigt profile fitting. 
\label{fig:ionfit} }
\end{center}
\end{figure*}

Physical properties of the CGM in galaxy formation simulations are sensitive to models of star formation and feedback \citep[e.g.,][]{Hummels2013, Suresh2015, FaucherGiguere2015, Liang2016}. For example, \cite{Liang2016} shows that metallicity of halo gas can change by a factor of 30 with a variation of star formation and feedback parameters. An advantage of combining synthetic spectrum generation tools with insights from the simulations is to make mock observations and directly compare quantities that are derived from them with the actual properties of the gas in the simulations. This can test the techniques used in observations, as well as make comparisons with simulations more meaningful. 

Without proper treatment of upper limits, observational analyses are often plagued by a limited number of detections of ions to constrain physical properties of the absorbing gas. 
Lower limits similarly prohibit precise constraints of metallicity, which usually occur in saturated Ly$\alpha$ line in the absence of Lyman series due to limited wavelength coverage. In this section, we present a Bayesian framework for deriving physical properties of the absorbing gas, specifically focusing on systems with poor constraints of column densities. Both saturated and undetected absorption lines make modelling the underlying physical properties very uncertain since they require some modelling of the probability distribution of the lower/upper limits themselves. 

Recently, a number of studies explored ways to maximize the utility of non-detections in observational analyses \citep[e.g.,][]{Crighton2015,Fumagalli2016} by modelling the corresponding posterior probability distribution function (PDF).  In particular, the PDF is assumed to be flat for column density below the upper limit and is joined by a one-sided Gaussian with constant standard deviation above the limit. However, the actual shape of the posterior distribution is determined by the nonlinear function through which parameters enter into a model of observables and, partly, through assumption about prior distributions. The latter may have a significant effect on the posterior distribution when the constraining power of the data is weak, which is exactly the regime of ``non-detections.'' As is well known, and as our results below illustrate, the posterior distribution in this regime can have a wide range of shapes, that are generally not consistent with the standard assumptions.

In our approach, we take advantage of knowing the underlying properties of the gas in the simulation and ability to generate synthetic spectra to illustrate how a Bayesian approach can be used to avoid making assumptions about the posterior distributions by directly constructing them from the data or making explicit and more justifiable assumptions through the priors based on plausible physical conditions at hand. In addition, a Bayesian approach allows a uniform treatment of both undetected and detected (including lower limits) absorption lines. In particular, the PDFs of detections need not be Gaussian since the degree of saturation will simply be reflected in the shape of the posterior PDF. 

Our pipeline starts with extraction of column densities of various ions using absorption lines in the spectrum and then uses the derived column density PDFs to constrain the density, temperature, and metallicity of the absorbing gas. In both steps, we use the Bayesian approach, combined with Monte Carlo Marko Chain (MCMC) sampling as we outline below. 

\subsection{Bayesian Voigt Profile Fitting}

We introduce a Bayesian approach to Voigt profile fitting in this section. We refer readers to our companion paper \citep{LiangKravtsov2017} where we describe in details the implementation and other useful utilities (e.g., continuum model) in a \texttt{PYTHON} package, \texttt{bayesvp}. The package is publicly available online\footnote{{\tt https://github.com/cameronliang/bayesvp}}.  

Briefly, the posterior distribution for fitting a Voigt profile ($\vec{\theta} = \{N, b, z\}$) to the spectrum (flux and its uncertainties; $\vec{F}$) can be written as:
\begin{equation} 
\pi_{\rm VP}(\vec{\theta} | \vec{F}) = \mathcal{L}_{\rm VP}(\vec{F} | \vec{\theta}) p_{\rm VP}(\vec{\theta}), 
\end{equation}
where $\pi_{\rm VP}(\vec{\theta} | \vec{F})$ is the product of the likelihood $\mathcal{L}_{\rm VP}(\vec{F} | \vec{\theta})$ with the prior of the parameters $p_{\rm VP}(\vec{\theta})$. The total likelihood is the product of all likelihoods given the individual (uncorrelated) velocity intervals of the spectrum: 

\begin{equation} \mathcal{L}_{\rm VP}(\vec{F} | \vec{\theta}) = \prod_{i} \ell_i(F_i | \vec{\theta})  \end{equation}

The likelihood for flux $F_i$ at a spectral pixel $i$, $\ell (F_i | \vec{\theta})$, represents the distribution of the fluctuations about the continuum, which can be measured in a spectral region free of absorption lines. For simplicity, it is usually assumed to be Gaussian (or Poisson distributed in photon limited cases).  As an example, if $\ell$ is a Gaussian, the form is: 
\begin{equation} \ell_i (F_i | \vec{\theta}) = \frac{1}{\sqrt{2 \pi \sigma_i^2}} \exp \left[-\frac{(F_i(\vec{\theta}) - F_i)^2}{2 \sigma_i^2} \right]\end{equation}

Here, we emphasize that the posterior $\pi_{\rm VP}(\vec{\theta})$ does not make any distinctions among detections, lower limits, and upper limits. By definition and design, the probability $\pi_{\rm VP}$ for a combination of $\{N, b, z\}$ is simply constrained by the spectral data. The posterior form is determined on the fly and is not assumed.  As mentioned above, we derive model parameters from MCMC chains using publicly available MCMC sampler \texttt{emcee} \footnote{\url{https://github.com/dfm/emcee}} \citep{GoodmanWeare2010,ForemanMackey2013} and \texttt{kombine} \footnote{\url{https://github.com/bfarr/kombine}} \citep{kombine}.  Note that the initializations of walkers do not affect the PDF as long as the chains have converged and the priors encompass the true values of model parameters.  For convergence criteria, we employ the Gelman-Rubin (GR) indicator \citep{GelmanRubin1992} with $\rm{GR} \leq 1.005$ \citep[see the implementation details in][]{LiangKravtsov2017}. In addition, we select the known number of Voigt profile components in our tests for simplicity. Nevertheless, the number of components (i.e the best model) can be identified using Bayesian evidence criteria implemented in our package \citep{LiangKravtsov2017}. 

We now turn our attention to example fits that bracket cases of non-detections and detections, including saturated lines. The top panel of Figure \ref{fig:vpfit} shows a very weak C\,II line buried inside a relatively noisy spectrum and the corresponding marginalized column density probability distribution given the data. Clearly, column density greater than $10^{13}\,\rm{cm^{-2}}$ would result in absorption larger than the noise level allows.  The posterior sampled by the MCMC chain disfavors larger values of column density. Therefore, the PDF for non-detection is flat for values $< 10^{13}\,\rm{cm^{-2}}$ and drops precipitously beyond that. 

The second and third rows of panels of Figure \ref{fig:vpfit} show when lines are detected but the derived column densities are uncertain. Most notably, the column density PDF of saturated H\,I line is bimodal due to the correlation between the Doppler parameter and the low $S/N$ of the spectra.  These two modes often are separated by two dex in column densities ($\sim 10^{15} \rm{cm^{-2}}$ and $\sim 10^{17} \rm{cm^{-2}}$).  Note that it is difficult for a traditional $\chi^2$ minimization grid search to capture the full degeneracy and locate both modes since it can be trapped in a $\chi^2$ local minimum.  This could result in a dramatic under- or overestimate of gas metallicity by orders of magnitudes if only one of these modes is adopted.  Finally, this bimodality reduces to a single mode with if $S/N$ of the spectrum is high enough. The bottom two panels show examples of well-detected absorption lines with Gaussian posteriors of the column density. 

\subsection{Bayesian Ionization Modelling}

\subsubsection{From the posteriors of column densities to posteriors of physical properties}

Given the posterior distributions of column density of low and intermediate ions derived for each ion from the spectra, we can use the Bayesian approach to constrain parameters of a single-phase ionization model. As an illustration, we apply such analysis for an ionization model that uses two basic assumptions: (1) the absorbing gas is ionized by a combination of photo and collisional ionization and is in ionization equilibrium and (2) absorbing gas can be modelled as a plane parallel slab described by a single density, temperature, and metallicity. These assumptions can, of course, be relaxed in more complicated models, which would, however, require more observational constraints in the form of more ion column density measurements. 

 For simplicity, we generate a grid of single-phase ionization models with parameters $ \vec{\alpha} = \{n_H, T, Z/Z_{\odot}, N_{\rm HI}\}$ using \texttt{CLOUDY} \citep{Ferland2013}. We assume the gas follows solar pattern. Note that we include $N_{\rm HI}$ as a parameter in the model \citep[see also][]{Crighton2015, Fumagalli2016},  such that $N_k = N_k(\vec{\alpha})$ is the column density of  the $k$th metal ion. Note also that the thickness of  the plane-parallel slab in the model is not a free parameter, as it is determined by a combination of model parameters, $\Delta S = \frac{N_{\rm HI}}{f_{\rm HI}(n_{\rm H}, T) n_{\rm H}}$, where $f_{\rm HI}(n_{\rm H}, T)$ is the ionization fraction of H\,I. 

The data constraining the model are a set of derived posterior distributions of column densities, $\vec{D} = \{\pi_{\rm{FeII}}(N), \pi_{\rm{SiII}}(N), \pi_{\rm{SiIII}}(N), ... \}$.  Specifically, $\pi_{\rm{k}}(N)$ is the probability distribution as a function of column density $N$ for metal ion $k$ marginalized over the $b$ parameter and redshift $z$. Recall that these posterior distributions are extracted using MCMC sampling, as described in section 3.1, and can be of any form depending on how well absorption lines of a given ion are detected.  We note that {\it all} co-spatial ionic species can be included in the ionization modelling, as long as there exists a transition within the wavelength coverage (and without contamination from systems of other redshifts). Within the framework described below the combination can also include detection of varying quality including upper and lower limits. We assume that the column density constraints for different ions are independent of each other. 

The posterior of the ionization model parameters $\vec{\alpha}$, given the column density probability distributions, can be written as: 
\begin{equation} 
\pi_{\rm ion}(\vec{\alpha} | \vec{D}) = \mathcal{L}_{\rm ion}(\vec{D} |\vec{\alpha} ) p(\vec{\alpha}),
\end{equation}
where the prior is simply: 
\begin{equation} p(\vec{\alpha}) = p(n_{\rm H}) p(T) p(Z/Z_{\odot}) \pi_{\rm HI}(N_{\rm HI}). \end{equation}
Note that $N_{\rm HI}$ is constrained using the measured probability distribution $\pi_{\rm HI}$ in the Voigt profile fitting process. We adopt uninformative priors for $\log n_{\rm H}$, $\log T$, and $\log  Z/Z_{\odot}$ such that all values are equally probable within a given range.  

The total likelihood $\mathcal{L}_{\rm ion}$ is the product of individual likelihood $\pi_k$ at $N_k$ for $k$ metal ions of a fixed set of parameters $\vec{\alpha}$:
\begin{equation} 
\mathcal{L}(\vec{D} | \vec{\alpha} ) =  \prod_{k} \pi_k (N_k(\vec{\alpha})).
 \end{equation}
In other words, if one measures the column density probability distributions of Si\,II, Si\,IV and Fe\,II, $\mathcal{L}_{\rm ion} = \pi_{\rm{FeII}}(N (\vec{\alpha})) \times \pi_{\rm{SiII}}(N (\vec{\alpha}) ) \times \pi_{\rm{SiIII}}(N (\vec{\alpha}) )$. In practice, we evaluate the probability for $\pi_k$ at a given $N_k$ through interpolated version of the MCMC sampled $\pi_k (N (\vec{\alpha}) )$. 

\subsubsection{Mock Observations and Experiments}

In the following, we conduct three sets of mock observations and experiments (A, B and C) to explore the relationship between the derived physical properties of absorbers and the underlying properties in the simulations through ionization modelling.  Briefly, Experiment A tests if we can recover the true properties over an entire LOS by including the total integrated column density despite the number of velocity absorption components. Experiment B tests if one can include higher ionization species with the low ions. Experiment C tests the recoverability of physical properties in poor column density constraints (i.e., with an abundance of lower limits). 

In experiment A, we broadly divide LOS into two types. The first set of LOS contains low and intermediate ions (i.e., HI, CII, CIII, CIV, SiII, SiIII, SiIV, OI, and FeII). The second type of LOS probes the coronal hot gas with a negligible abundance of low metal ions. We, therefore, include H\,I and high ions only (e.g., N\,V, O\,VI, and Ne\,VIII). Since it is often difficult to distinguish the spatial information along the LOS based on velocity information in real spectra, we are interested in testing whether the LOS H\,I - weighted properties can be recovered despite the number of velocity/absorption components. Therefore, we simply test the recoverability of the physical properties during the ionization modelling process by assuming the column densities are well measured with uncertainties of $\sigma_{\log N} =$ 0.1 dex for all ions. We do so by directly integrating the number density in our simulations in this exercise. This implies that we include all absorption/velocity components in the spectra.  Nevertheless, in Figure \ref{fig:props_comp}, we find that the derived values generally follow the HI mass-weighted properties along the LOS.  Occasionally, catastrophic failures occur when there is more than one density peak are present in the LOS (shown by data points with small error bar but large deviation from the truth values). This is because the ratio of the sum of metals and H\,I is not equivalent to the ratio of metal and H\,I in each density peak. In other words, if these structures overlap in velocity space where column density contribution cannot be separated, single phase ionization model would fail to capture properties of both structures. It is also worth pointing out the deviation at low-temperature regime (near $10^4$ K) in Figure \ref{fig:props_comp} is due to the prior that prohibits MCMC walkers to explore temperature lower than $10^4$ K. The posterior distribution is therefore asymmetric; the median is artificially offset despite having the most probable temperature at $10^4$ K. This can systematic shift can be fixed by including CLOUDY models below $10^4$ K and relaxing the priors. 

It is common to only include ions that are co-spatial to extract properties using a single-phase model. This is because a single-phase model predicts sensible properties only if ions arise from the same density and temperature. For experiment B, in contrast, we test explicitly what properties one would derive if we include column densities of ions which are not co-spatial. Note that we only consider isolated peaks in contrast to the discussion of multiple peaks above. We do so by extracting small spatial segments (typically few kpc across) of LOS containing a single structure with varying density and temperature (such as the density peak in the left panel of Figure \ref{fig:mp_spat}). For each density structure, we perform a series of ionization fit by cumulatively including column densities of ions with increasing ionization energy, therefore tracing more and more diffuse gas at the interface of the dense absorber and the halo. For example, we first include 4 low ions with $E_{\rm ion} \le 16.345$ eV (i.e H\,I, O\,I, Mg\,II and Si\,II). Next, we add the ion with higher ionization potential (CII with $E_{\rm ion} = 24.383$ eV) to test how the derived properties change, and so on up to O\,VI with $E_{\rm ion} = 138.11$ eV. Somewhat surprisingly, Figure \ref{fig:ionfitB} shows that by including ions with increasing $E_{\rm ion}$, the derived properties all converge to the H\,I mass-weighted properties over the density structure similar to experiment A. An example of a corner plot that includes both low and high ions are shown in Figure \ref{fig:ionfitB3}, where the recovered parameters are consistent with the underlying LOS values.  Given that the derived properties are HI-mass weighted along a peak, the derived values slowly change depending on the ions included in the analysis.  In other words, including low ions only in the modeling will give density, temperature, and metallicity closest to the highest density region. Including additional high ions will drive the derived density lower since high ions occur in the outskirt of the structure where density is lower. 

Note that our finding here is in contrast to our usual expectation where O\,VI should not be included in the ionization modelling \citep[e.g.,][]{Churchill2015}. It is conceivable that what we show here can break down in higher resolution simulations where the morphology can be much more complex.  It is also important to note that we use the same underlying ionization model that produce the ions abundances in the inverting process. In reality, we might not have full knowledge of the ionization process in observed ionic abundances which can jeopardize the retrieval of absorbers physical properties. Other sources in addition to the UVB may provide additional ionizing photons that change the relative abundance of ions \citep[e.g. the highly ionized gas O\,VI discussed in][]{Werk2016}. 

Finally, we design experiment C to illustrate the uniform treatment of detections and non-detections. To avoid possible confusion from experiment A and B, we generate an idealized single phase model with a random combination of density, temperature, metallicity and velocity dispersion. A noise-free synthetic spectrum can then be created to represent such isothermal clouds. We then add noise to the synthetic spectra to the level (S/N = 20) that weak lines are not detectable in the traditional sense (i.e., by eye). The primary motivation for doing so is to test the utilities of upper limits. This is because detections in observed spectra can be scarce.  It is especially useful to include proper treatment of upper limits when the number of detections is less than the ionization model parameters.  Figure \ref{fig:props_comp} shows ionization modelling results for 30 LOS with varying quality of column density constraints. In some cases where there is a mix of detections and non-detections, the posteriors are similar to that shown in Figure \ref{fig:ionfit} when constraints are poor and are similar to Figure \ref{fig:ionfitB3} when constraints are good. 

The corner plot of Figure \ref{fig:ionfit} shows the posterior distributions of density, temperature, and metallicity of the hot halo gas. Here, we use ``detected" ions (H\,I, O\,VI, and N\, V) and one upper limits constraint (Ne\,VIII) as input for a four-parameter model. Despite the poor constraint in Ne\,VIII column density,  the temperature and metallicity of the gas are recovered. Although constraint on the density is poor, this test shows that one can formally recover the true parameters to the degree that the data allow. The large error bars of the derived properties from these coronal LOS illustrate that the absorption spectrum, in this case, has little constraining power. Explicit priors can also be set to limit the any unphysical values of model parameters based on our understanding of the absorbing gas physical condition (e.g., unphysically large density for O\,VI-producing gas). 

\section{Conclusions}

We explored the multiphase structure of the CGM probed by synthetic spectra through a cosmological zoom-in galaxy formation simulation. We use simulations to produce synthetic absorption spectra containing lines of different ionic species. In addition, we present a Bayesian method for modelling a combination of absorption lines to derive physical properties of absorbers regardless of the signal-to-noise or whether a particular line is actually detected. In other words, the method treats detections (including lower limits) and non-detections (upper limits) in a uniform manner.  Our main findings can be summarized as follows. 

\begin{enumerate}

    \item[1.] We show that in our simulations low ions (e.g., H\,I, Mg\,II), arise from the densest and coldest regions of absorbers, while intermediate and high ions arise in the interface regions between cold gas and hot halo, as shown in Figure \ref{fig:mp_spat} \citep[see also][]{Churchill2015}. 
    
    \item[2.] In the lines of sight passing through the cold gas, absorption lines of low, intermediate and high ions are present in the spectrum and overlap in velocity space. This is because the velocity across a typical absorber does not vary significantly over sub-kpc scale in our simulations. In lines of sight that passing the hot halo, only H\,I and high ions are present with significant column densities. In particular, we find detectable O\,VI and O\,VII at $\log N \sim 13-14,\rm{cm^{-2}}$ in our simulations. In such lines of sight, the absorption lines are typically broad due to the complex velocity fields across the entire halo \cite[Figure \ref{fig:mp_spat} and Figure \ref{fig:mp_spec}; see also][]{Savage2011a, Werk2016}. 
    
    \item[3.] We find that attenuation of the UVB flux by neutral hydrogen in the CGM increases the abundance of low ions by a factor of a few to an order of magnitude. This difference, however, is smaller than LOS to LOS scatters. A comparison between photo- and collisional ionization models show that low and intermediate ions in our simulations are mostly photoionized, while high ions (e.g., OVI) is primarily collisionally ionized. In addition, photoionization from the UVB becomes more important at the outskirts of the halo ($>0.25 R_{\rm vir}$) for low ions due to the decreasing covering fraction of dense gas. The decreasing density and constant UVB together are responsible for shaping the exponential form of the column density profiles of low ions (Figure \ref{fig:spec_ion_model} and Figure \ref{fig:ionpro}). 

    \item[4.] We present a Bayesian method for Voigt profile fitting that treats detections and non-detections (upper and lower limits) in a uniform manner and produces posterior of column densities, b-parameters, and redshifts, reflecting the quality of the absorption lines.  We also use Bayesian approach to derive density, temperature, and metallicity of absorbers from using these posteriors. 

    \item[5.] We show that the derived properties match closely the corresponding H\,I mass-weighted averages along the LOS. While this works well for an isolated structure with internal varying density, it also implies that single-phase model would fail catastrophically in multiple density peaks, especially if they overlap in velocity space such that their column density contribution are not well separated.  Finally,  we also show that constraints on H\,I and O\,VI together with upper limits on Ne\,VIII or N\,V, can in principle provide constraints on the properties of the hot halo (Figure \ref{fig:vpfit} - \ref{fig:ionfit}).

\end{enumerate}

The methods developed in this study can be used to make comparisons between simulations and observations more realistic using synthetic spectra. The results of our analysis indicate density, temperature, and metallicity of the CGM absorbing gas can be reliably recovered in a Bayesian ionization modelling presented here. 


\section*{Acknowledgments}

CL thanks Sean Johnson for his helpful comments. We also thank the anonymous referee for helpful comments and improving the presentation of the manuscript. CL and AK were supported by a NASA ATP grant NNH12ZDA001N, NSF grant AST-1412107, and by the Kavli Institute for Cosmological Physics at the University of Chicago through grant PHY-1125897 and an endowment from the Kavli Foundation and its founder Fred Kavli.  OA acknowledges support from the Swedish Research Council (grant 2014-5791) and the Knut and Alice Wallenberg Foundation. CL is partially supported by NASA Headquarters under the NASA Earth and Space Science Fellowship Program - Grant NNX15AR86H. The simulations presented in this paper have been carried using the Midway cluster at the University of Chicago Research Computing Center, which we acknowledge for support.






\bibliographystyle{mnras}
\bibliography{ms} 








\bsp	
\label{lastpage}
\end{document}